\documentclass[
    reprint,
    onecolumn,
    notitlepage,
    aps,
    prd,
    11pt,
    superscriptaddress,
    longbibliography
]{revtex4-1}

\usepackage{amsmath,amssymb}
\usepackage{graphicx}
\usepackage{placeins}
\usepackage{microtype}
\usepackage{xurl}
\usepackage[dvipsnames]{xcolor}
\usepackage[colorlinks=true,linkcolor=Blue,citecolor=Blue,urlcolor=Blue,hypertexnames=false]{hyperref}

\usepackage[nameinlink,noabbrev]{cleveref}
\usepackage[acronym,nomain,nonumberlist]{glossaries-extra}
\setabbreviationstyle[acronym]{long-short}
\glsdisablehyper

\newacronym{md}{MD}{molecular dynamics}
\newacronym{gpu}{GPU}{graphics processing unit}
\newacronym{gpumd}{GPUMD}{Graphics Processing Units Molecular Dynamics}
\newacronym{wpe}{WPE}{Wisevolve Potential Engine}
\newacronym{nep}{NEP}{neuroevolution potential}
\newacronym{qnep}{qNEP}{neuroevolution potential with dynamic charges}
\newacronym{pimd}{PIMD}{path-integral molecular dynamics}
\newacronym{lj}{LJ}{Lennard-Jones}
\newacronym{eam}{EAM}{embedded-atom method}
\newacronym{bec}{BEC}{Born effective charge}
\newacronym{pppm}{PPPM}{particle-particle-particle-mesh}
\newacronym{rdf}{RDF}{radial distribution function}
\newacronym{hnemd}{HNEMD}{homogeneous nonequilibrium molecular dynamics}
\newacronym{nvt}{NVT}{canonical ensemble}
\newacronym{npt}{NPT}{isothermal-isobaric ensemble}
\newacronym{abi}{ABI}{application binary interface}

\newcommand{\gpumdwpe}{GPUMD-WPE}

\begin{document}
\title{Wisevolve Potential Engine: Machine-Learning Molecular Dynamics at 100 Million Atom-Steps per Second on a Single GPU}

\author{Hekai Bu}
\email{hekai\_bu@whu.edu.cn}
\affiliation{Wisevolve, Jinzhou, P. R. China}
\affiliation{College of Physical Science and Technology, Bohai University, Jinzhou 121013, P. R. China}

\author{Zheyong Fan}
\email{brucenju@gmail.com}
\affiliation{Wisevolve, Jinzhou, P. R. China}
\affiliation{College of Physical Science and Technology, Bohai University, Jinzhou 121013, P. R. China}

\begin{abstract}
General-purpose molecular dynamics software supports diverse interatomic potentials and simulation requirements.
Within a simulation stage, the loaded potential model and the required calculations typically remain fixed as the atomic configuration evolves.
We present the Wisevolve Potential Engine (WPE), which uses this information from GPUMD to select and reuse a GPU execution policy tailored to the model and simulation stage.
GPUMD retains simulation control and ownership of the simulation data.
On a single NVIDIA RTX 5090, WPE-enabled GPUMD (GPUMD-WPE) achieves a median throughput of 109 million atom-step/s across 106 neuroevolution potential (NEP) workloads. 
For a 15.1-million-atom diamond Si system, it sustains 142 million atom-step/s. Relative to GPUMD, geometric-mean speedups are 3.49$\times$ across these NEP workloads,
3.61$\times$ across 16 workloads using NEP with dynamic charges (qNEP), 1.24$\times$ across 28 classical-potential workloads, and 10.5$\times$ across 56 path-integral molecular dynamics (PIMD) configurations.
End-to-end calculations of water structure, lattice thermal conductivity, and PIMD thermal expansion show close agreement with corresponding GPUMD results and published simulation results.
A separate 12.5-million-atom Cu shock-compression simulation demonstrates its applicability to large-scale nonequilibrium dynamics.
These results support model- and stage-specific specialization as a practical way to accelerate GPU potential evaluation in general-purpose molecular dynamics workflows.
\end{abstract}
\maketitle

\glsresetall

\section{Introduction}
\label{sec:introduction}

\Gls{md} probes structural, thermodynamic, mechanical, and transport properties of materials by simulating atomic motion.
Repeated interatomic force evaluations often account for a substantial fraction of the computational cost. Much of this work can run concurrently over atoms and their local environments, making MD well suited to \gls{gpu} acceleration.
\Gls{gpumd} is a general-purpose GPU-based MD package for large-scale atomistic simulations \cite{Fan2022GPUMD,Xu2025GPUMD4}. Its applications span thermal transport, mechanical response, radiation damage, phase transitions, shock compression, and ion transport across a wide range of materials \cite{Xu2025GPUMD4}.

\gls{gpumd} supports these applications with diverse interatomic potentials and simulation methods.
Its empirical potentials, including \gls{lj}, \gls{eam}, and Tersoff, describe atomic interactions through prescribed analytical forms \cite{Jones1924LJ,Daw1984EAM,Tersoff1989,Xu2025GPUMD4}.
For more flexible interaction models, \gls{gpumd} provides the native \gls{nep} framework, combining local structural descriptors with a compact neural network for efficient \gls{gpu} evaluation \cite{Fan2021NEP,Fan2022GPUMD}.
Within this framework, general-purpose models such as UNEP-v1 \cite{Song2024UNEP16} and NEP89 \cite{Liang2026NEP89} broaden the chemical coverage.
To treat charge redistribution and long-range electrostatics explicitly, the \gls{qnep} extension introduces environment-dependent charges and an electrostatic contribution to the interaction energy \cite{Fan2026qNEP}.
\gls{gpumd} also supports \gls{pimd} for nuclear quantum effects \cite{Ying2025PIMD,Xu2025GPUMD4}.

These models and simulation methods require GPUMD to support potential-evaluation workloads that differ in potential family, model shape, and simulation requirements.
Within a simulation stage, however, the loaded model and required computations typically remain fixed while the atomic configuration evolves.
This stability allows a specialized execution policy to be selected and reused across GPU potential evaluations.

We present the \gls{wpe}, which specializes GPU potential evaluation within GPUMD using the loaded model and the simulation-stage requirements.
We benchmark GPUMD-WPE across multiple potential families and simulation methods and validate its simulation results through representative end-to-end calculations. 
To the best of our knowledge, this is the first reported demonstration of MD with machine-learned interatomic potentials at over 100 million atom-step/s on a single  GPU. On an NVIDIA RTX 5090, \gpumdwpe{} sustains 142 million atom-step/s for a 15.1-million-atom diamond Si system.
A 12.5-million-atom Cu shock-compression simulation also demonstrates the use of \gpumdwpe{} in large-scale nonequilibrium dynamics.

\section{Background and Wisevolve Potential Engine Framework}
\label{sec:background}
\label{sec:design}

\subsection{Interatomic Potentials and Path-Integral Molecular Dynamics}
\label{sec:background:potentials}

For a total potential energy $U$, the conservative force on atom $i$ is
\begin{equation}
    \mathbf F_i=-\nabla_i U(\{\mathbf r\}),
\end{equation}
where $\{\mathbf r\}$ denotes the atomic configuration.
Despite this common force-energy relation, different representations of $U$ require distinct operations and dependencies when evaluating energies, forces, and virials.

Classical empirical potentials describe atomic interactions through prescribed functional forms with parameters fitted to reference data. Writing the total energy as $U=\sum_i U_i$, a simple example is the \gls{lj} potential \cite{Jones1924LJ}, whose site energy depends only on pairwise separations,
\begin{equation}
    U_i^{\mathrm{LJ}} =
    \frac{1}{2}\sum_{j\ne i}4\epsilon_{IJ}
    \left[
    \left(\frac{\sigma_{IJ}}{r_{ij}}\right)^{12}
    -
    \left(\frac{\sigma_{IJ}}{r_{ij}}\right)^6
    \right],
\end{equation}
where $r_{ij}$ is the distance between atoms $i$ and $j$, $I$ and $J$ denote their species, and $\epsilon_{IJ}$ and $\sigma_{IJ}$ set the energy and length scales of the interaction. Each pair contribution can therefore be evaluated directly from the interatomic distance. Incorporating the local environment introduces additional computational dependencies. In the \gls{eam} \cite{Daw1984EAM}, developed for metallic systems, the site energy combines a pair contribution with an embedding energy that depends on the local electron density,
\begin{equation}
    U_i^{\mathrm{EAM}}=
    \frac{1}{2}\sum_{j\ne i}\phi_{IJ}(r_{ij})+F_I(\rho_i),
    \qquad
    \rho_i=\sum_{j\ne i}f_J(r_{ij}),
\end{equation}
where $\phi_{IJ}$ is the pair potential, $F_I$ is the embedding-energy function, and $f_J$ gives the density contribution from a neighboring atom of species $J$. The density must be accumulated before the embedding energy and its derivative can be evaluated, so force evaluation depends on density accumulation. Within this general framework, EAM variants differ in the functional forms of the pair interaction, density contribution, and embedding energy. Examples include the Zhou-2004 EAM formulation \cite{Zhou2004EAM} and the Dai-2006 extended Finnis-Sinclair formulation \cite{Dai2006EAM}. For covalent systems, bond-order potentials account for the surrounding bonding geometry. In the Tersoff-1989 potential used here \cite{Tersoff1989}, the site energy is
\begin{equation}
    U_i^{\mathrm{T89}}=
    \frac{1}{2}\sum_{j\ne i}f_C(r_{ij})
    \left[f_R(r_{ij})-b_{ij}f_A(r_{ij})\right],
\end{equation}
where $f_C$ is a cutoff function, $f_R$ and $f_A$ are repulsive and attractive radial terms, and $b_{ij}$ modulates the attractive interaction according to the local environment,
\begin{equation}
    b_{ij}=
    \chi_{IJ}
    \left(1+\beta_I^{n_I}\zeta_{ij}^{n_I}\right)^{-1/(2n_I)},
    \qquad
    \zeta_{ij}=
    \sum_{k\ne i,j}f_C(r_{ik})g_{ijk}.
\end{equation}
Here, $g_{ijk}$ describes the dependence on the bond angle $\theta_{ijk}$, while $\chi_{IJ}$, $\beta_I$, and $n_I$ are model parameters. Evaluating each bond therefore requires an additional sum over surrounding atoms. These examples illustrate how pairwise interactions, density-dependent embedding, and bond-order modulation lead to distinct neighborhood calculations and data dependencies.

Unlike empirical potentials with prescribed analytical forms, machine-learned interatomic potentials use flexible function approximators, particularly neural networks, to capture complex nonlinear relationships between atomic environments and energies. Within \gls{gpumd}, the \gls{nep} framework \cite{Fan2021NEP,Fan2022GPUMD} implements this approach by encoding the local environment of each atom in a descriptor vector and mapping it to a site energy using a compact neural network. The total energy is 
\begin{equation}
    U=\sum_i U_i,
    \qquad
    U_i=\mathcal N_{\theta}(\mathbf q_i),
\end{equation}
where $\mathbf q_i$ is the descriptor vector of atom $i$ and $\theta$ denotes the neural-network parameters. The descriptor combines radial components that encode neighbor distances with angular components that capture correlations among neighboring bonds. Representative radial and three-body angular components are
\begin{align}
    q_n^i &= \sum_{j\ne i}g_n^{\mathrm R}(r_{ij}), \\
    q_{nl}^i &= \frac{2l+1}{4\pi}
    \sum_{j\ne i}\sum_{k\ne i}
    g_n^{\mathrm A}(r_{ij})g_n^{\mathrm A}(r_{ik})
    P_l(\cos\theta_{ijk}),
\end{align}
where $g_n^{\mathrm R}$ and $g_n^{\mathrm A}$ are radial functions constructed from basis expansions with learned coefficients and separately specified cutoff distances; their species dependence is left implicit. Here, $P_l$ is the Legendre polynomial of order $l$, and $\theta_{ijk}$ is the angle between bonds $ij$ and $ik$. Four-body and five-body angular components can also be included to capture higher-order correlations in the local environment \cite{Fan2022GPUMD}. The species count, radial and angular basis sizes, descriptor orders, included body-order components, cutoff distances, and neural-network width govern the dimensions of descriptor and network evaluation, the range of local neighborhoods, and intermediate-storage requirements. These properties define the model shape.

Standard \gls{nep} describes atomic interactions through local environments within finite cutoffs. The \gls{qnep} framework \cite{Fan2026qNEP} adds explicit long-range electrostatics through an environment-dependent partial charge as a second neural-network output,
\begin{equation}
    (U_i,Q_i)=\mathcal N_{\theta}(\mathbf q_i),
\end{equation}
where $U_i$ and $Q_i$ denote the short-range site energy and predicted partial charge, respectively. A total-charge correction enforces charge conservation before electrostatic evaluation. The total energy combines the local NEP and electrostatic contributions,
\begin{equation}
    U^{\mathrm{tot}}
    =
    U^{\mathrm{NEP}}
    +
    U^{\mathrm{ES}},
    \qquad
    U^{\mathrm{NEP}}=\sum_i U_i.
\end{equation}
Under periodic boundary conditions, qNEP provides two formulations of $U^{\mathrm{ES}}$ that differ in their explicit treatment of short-range electrostatics:
\begin{equation}
    U^{\mathrm{ES}}=
    \begin{cases}
        U^{\mathrm r}+U^{\mathrm k}+U^{\mathrm s},
        & \text{Mode~1},\\
        U^{\mathrm k},
        & \text{Mode~2},
    \end{cases}
\end{equation}
where $U^{\mathrm r}$ and $U^{\mathrm k}$ are the real- and reciprocal-space contributions, respectively, and $U^{\mathrm s}$ is the self-energy correction. Mode~2 leaves short-range interactions to the local NEP term and includes only the reciprocal-space contribution explicitly. In both modes, $U^{\mathrm k}$ can be evaluated using direct Ewald summation or the \gls{pppm} method. The charges also provide access to polarization and \gls{bec} tensors \cite{Fan2026qNEP}. Thus, the electrostatic formulation, reciprocal-space solver, and requested charge-related quantities also affect potential evaluation requirements.

While the potentials above specify interatomic interactions, accounting for nuclear quantum effects additionally requires a quantum statistical treatment of the nuclei. In \gls{pimd}, each nucleus is represented as a ring polymer of $P$ beads connected by harmonic springs \cite{Ying2025PIMD}. For a system of $N$ nuclei, each bead index defines a configuration containing one bead from each nucleus, and the same interatomic potential $U$ is applied to every such configuration. The physical-potential contribution to the ring-polymer Hamiltonian is therefore
\begin{equation}
    U_P=
    \sum_{s=1}^{P}
    U\!\left(
    \mathbf r_s^{(1)},\ldots,\mathbf r_s^{(N)}
    \right),
\end{equation}
where $\mathbf r_s^{(i)}$ denotes the position of bead $s$ of nucleus $i$. Each force step thus requires $P$ evaluations of the same potential, one for each bead configuration. These evaluations share the model parameters but retain separate coordinates, forces, energies, and virials for each bead. The bead count consequently determines the extent of repeated work in potential evaluation.

The potentials and simulation methods discussed above may also be used for systems containing only a few atoms in small periodic cells. When the cell is small relative to the interaction cutoff, an atom's neighborhood may contain multiple periodic images of the same atom. In sufficiently small cells, the central atom's own images can also lie within the cutoff. All such image contributions must be included in potential evaluation. At the same time, small systems typically require much less work per evaluation than large-scale simulations and offer limited atom-level parallelism. Execution schemes designed for large systems may therefore underutilize the \gls{gpu}. 

These requirements can occur in combination.
For example, \gls{qnep} in \gls{pimd} requires both local and electrostatic evaluation for each bead configuration, while small periodic cells require periodic-image handling in either conventional MD or \gls{pimd}.

\subsection{WPE Framework}
\label{sec:design:framework}

\begin{figure}[!htb]
    \centering
    \includegraphics[width=0.98\linewidth]{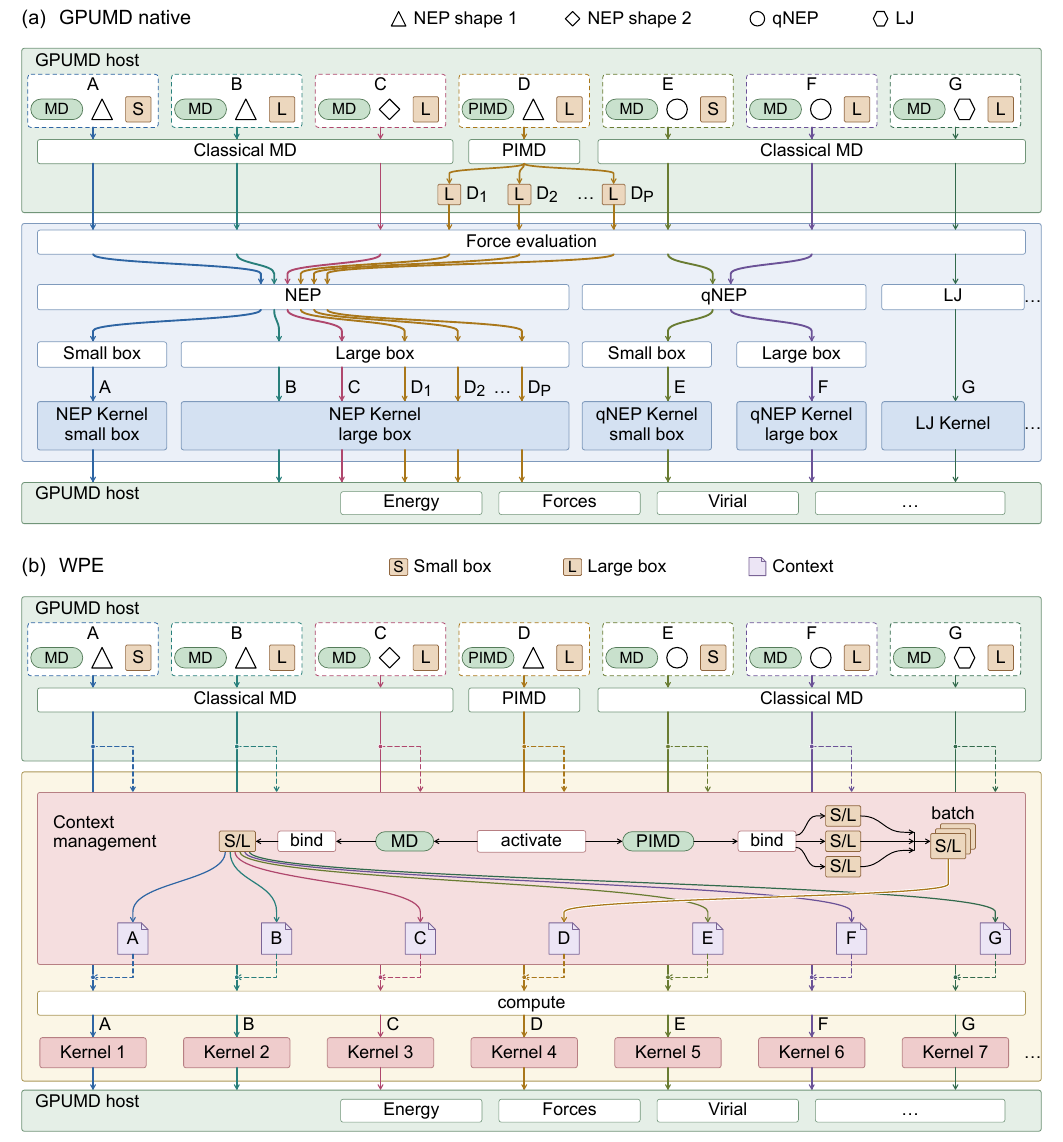}
    \caption{Native GPUMD and WPE potential-evaluation workflows. Colored routes trace workloads A–G. (a) Native GPUMD dispatches by potential family and box type; different model shapes can share an evaluation path, and PIMD beads are evaluated separately. (b) WPE selects and reuses execution policies through persistent contexts containing potential family, model shape, stage requirements, and execution structure. Compatible PIMD bead configurations can be evaluated in batches. GPUMD retains simulation control and data ownership.}
    \label{fig:architecture}
\end{figure}

WPE performs potential evaluation, while GPUMD handles input processing, time integration, ensemble control, and observable scheduling, and retains ownership of the simulation data. GPUMD supplies the model and stage requirements that WPE uses to select an execution policy.
\Cref{fig:architecture} compares native GPUMD's dispatch by potential family and box type with WPE's selection and reuse of execution policies through persistent contexts, including batched evaluation of compatible \gls{pimd} bead configurations.

We implemented the interface between GPUMD and WPE through a versioned pure-C \gls{abi} with an opaque backend context. 
The context is associated with the loaded potential model. As GPUMD processes \texttt{run.in} sequentially, the integration layer collects the settings needed to describe the current stage. When GPUMD reaches an execution command such as \texttt{run} or \texttt{minimize}, this layer assembles the computational requirements for that stage and passes them to WPE through \texttt{activate}. The same model context can be retained when \texttt{activate} is called again for a subsequent stage.
During an active stage, \texttt{bind} supplies box information and references to GPUMD-owned device arrays, while \texttt{compute} performs potential evaluation using the current binding and selected execution policy.
An unchanged binding can be reused across repeated \texttt{compute} calls. The context is released through \texttt{destroy} when the potential is destroyed.

Native GPUMD routes potential evaluation by potential family and box type, as reflected in the kernel labels in \Cref{fig:architecture}(a). For example, workloads B and C share the NEP large-box implementation, with their different model dimensions handled through runtime parameters.
In panel (b), the numbered WPE kernels represent specialized implementations selected for the corresponding workload contexts. Each kernel box may involve multiple \gls{gpu} kernel launches.
WPE uses this interface to select an execution policy based on four dimensions: potential family, model shape, stage requirements, and execution structure. The potential family determines the evaluation path for \gls{nep}, \gls{qnep}, \gls{lj}, \gls{eam}, or Tersoff-1989. Model shape specifies the fixed computational dimensions of the loaded model. WPE uses these dimensions to select an execution path and organize reusable working memory, while atomic coordinates and neighbor-list state remain dynamic. Stage requirements specify the outputs and, for \gls{qnep}, the active electrostatic treatment and requested charge-related quantities. Execution structure describes how the work is organized across atoms and repeated evaluations, accounting for system size, periodic-image handling in small cells, and the bead dimension in \gls{pimd}. 
In eligible PIMD stages, WPE batches potential evaluations for multiple bead configurations.
Model parameters are shared, while the data for each bead remain separate.

WPE currently provides execution paths for supported \gls{nep}, \gls{qnep}, \gls{lj}, \gls{eam}, and Tersoff-1989 models, including the \gls{pimd} configurations evaluated here.
Simulation and analysis functions can be used when WPE supports their potential-evaluation requirements.
Other GPUMD workloads can use its native potential implementations. WPE handles kernel selection and hardware-specific tuning internally.

\section{Performance Evaluation}
\label{sec:benchmark}

\subsection{Benchmark Design and Overview}
\label{sec:benchmark:method}

In this section, we evaluate \gpumdwpe{} against \gls{gpumd} v5.9 on a single NVIDIA RTX 5090 GPU with CUDA 12.8.
Both implementations use mixed precision, with single precision for potential calculations and double precision for all other calculations. The qNEP benchmarks use PPPM for reciprocal-space electrostatics.
The evaluation includes four paired benchmark suites covering ordinary \gls{nep}, \gls{qnep}, classical potentials, and \gls{pimd}, together with a water-NEP system-size scaling study and a cross-family throughput survey.
The non-PIMD suites include both \gls{nvt} and \gls{npt} workloads, whereas the PIMD suite uses NVT-PIMD.
Throughput is calculated as the number of atom-steps divided by the total MD production time, including potential evaluation, time integration, and ensemble control. Both implementations use identical inputs for each paired workload, with three repeated measurements per implementation. The reported throughput is the arithmetic mean of the three measurements.
All benchmarks use 20 warm-up steps before production timing and a time step of 1~fs,
except for water systems, for which a time step of 0.5~fs is used.
\Cref{tab:benchmark-suite} summarizes the benchmark systems, atom counts, and production lengths.
For PIMD, production lengths vary with system size and bead count, with measured production times exceeding 2~s.

\begin{table}[!htb]
\centering
\renewcommand{\arraystretch}{1.22}
\small
\caption{Benchmark systems, atom counts, and production lengths.
All step counts exclude warm-up.}
\label{tab:benchmark-suite}

\begin{tabular}{
    @{}p{0.15\linewidth}
       p{0.12\linewidth}
       p{0.27\linewidth}
       p{0.27\linewidth}
       p{0.12\linewidth}@{}}
\hline
Suite & Workloads & Systems / cases & Atoms & Steps \\
\hline

Ordinary NEP
& 106
& 53 model-system cases
& 331,776--1,221,240
& 500 \\

NEP scaling
& 44
& Water
& 81--11,490,336
& 300--50,000 \\

qNEP
& 16
& Water; BaTiO$_3$
& 50,400; 53,240
& 5000 \\

LJ
& 6
& Ar; Kr; Xe
& 442,368
& 5000 \\

EAM
& 18
& 9 material-potential cases
& 432,000--442,368
& 5000 \\

Tersoff-1989
& 4
& Si; SiGe
& 512,000
& 5000 \\

PIMD
& 56
& Water
& 756--41,472 physical atoms
& 500--5000 \\

\hline
\end{tabular}
\end{table}

Let $S_m$ denote the speedup for paired workload $m$, defined as the ratio of the mean throughput of GPUMD-WPE to that of GPUMD.
The overall speedup across a benchmark suite containing $M$ workloads is reported as the geometric mean
\begin{equation}
    G=\exp\!\left(
        \frac{1}{M}\sum_{m=1}^{M}\ln S_m
    \right),
    \label{eq:gmean}
\end{equation}
which gives equal weight to the multiplicative speedup of each workload \cite{Hoefler2015Benchmarking}.
For PIMD, throughput is normalized by the number of physical atoms without multiplying by the bead count.

\begin{figure}[!htb]
    \centering
    \includegraphics[width=0.95\linewidth]{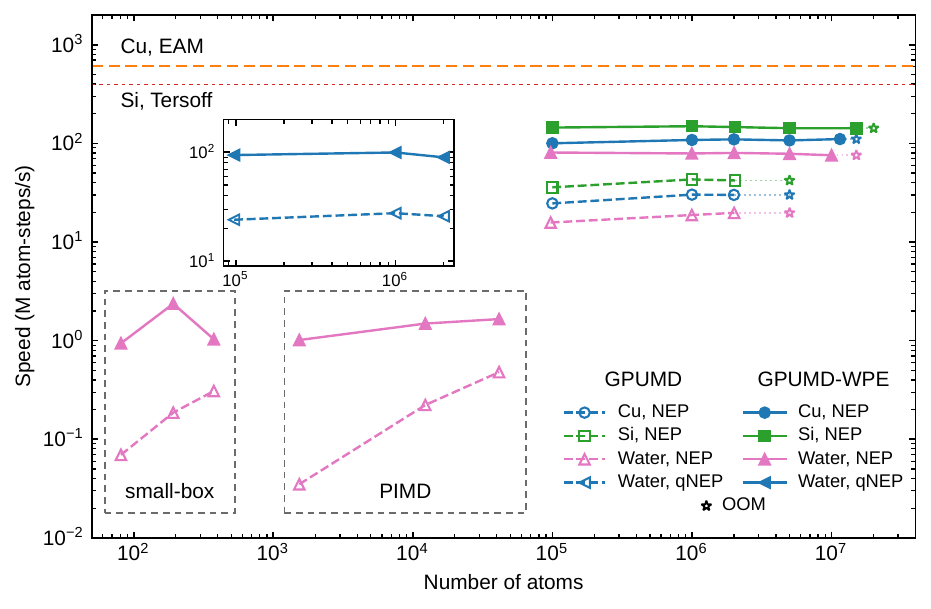}
    \caption{Absolute throughput across potential families, system sizes, and execution modes on a single NVIDIA RTX 5090.
Solid lines and filled symbols represent GPUMD-WPE, while dashed lines and open symbols represent GPUMD.
Dashed boxes highlight small-box water NEP and 32-bead water PIMD, and the inset shows water qNEP.
Horizontal lines indicate Cu EAM and Si Tersoff reference measurements; stars mark out-of-memory (OOM) runs. OOM markers are placed at the throughput of the preceding successful run for visualization only and the connecting dotted segments do not represent measured throughput.}
    \label{fig:overall-scaling}
\end{figure}

We complement the paired  benchmarks with an absolute-throughput survey across potential families, system sizes, and execution modes, shown in \Cref{fig:overall-scaling}.
The survey covers representative NVT simulations of Cu and Si with NEP, water with NEP and qNEP, Cu with EAM, and Si with Tersoff.
The Cu, Si, and water NEP cases use the NEP89 model \cite{Liang2026NEP89}, while the water qNEP case uses the model reported by Fan et al. \cite{Fan2026qNEP}.
It also includes small-box water NEP systems containing 81--375 atoms and 32-bead water PIMD.
Each throughput is based on three production runs, with lengths adjusted to system size.

For the large-scale NEP workloads, \gpumdwpe{} achieves throughputs of 110, 142, and 75.7 million atom-step/s for Cu, Si, and water systems containing 11.5, 15.1, and 10.0 million atoms, respectively.
The horizontal lines show GPUMD-WPE reference throughputs of 605 and 395 million atom-step/s for Cu EAM and Si Tersoff at approximately five million atoms, rather than scaling trends.
The out-of-memory (OOM) markers denote failed runs at selected  sizes, not definitive memory-capacity limits.
These survey results are not included in the paired-suite geometric means.

The four paired benchmark suites comprise 206 workloads.
Their geometric-mean speedups and ranges are summarized in \Cref{tab:performance-summary}.

\FloatBarrier

\begin{table}[!htb]
\centering
\renewcommand{\arraystretch}{1.22}
\caption{Performance summary for GPUMD-WPE relative to GPUMD on a single
NVIDIA RTX 5090. Geometric means and ranges are calculated over the workloads
in each paired suite.}
\label{tab:performance-summary}
\begin{tabular*}{0.76\linewidth}{@{\extracolsep{\fill}}lrrr@{}}
\hline
Workload family & Workloads & Geometric mean & Speedup range \\
\hline
NEP       & 106 & 3.49$\times$  & 2.36--5.62$\times$ \\
qNEP      & 16  & 3.61$\times$  & 2.70--4.82$\times$ \\
Classical & 28  & 1.24$\times$  & 1.02--2.18$\times$ \\
PIMD      & 56  & 10.5$\times$ & 2.75--37.7$\times$ \\
\hline
\end{tabular*}
\end{table}

\FloatBarrier
\subsection{NEP Performance and Scaling}
\label{sec:benchmark:nep}

\begin{figure}[!htb]
    \centering
    \includegraphics[width=0.95\linewidth]{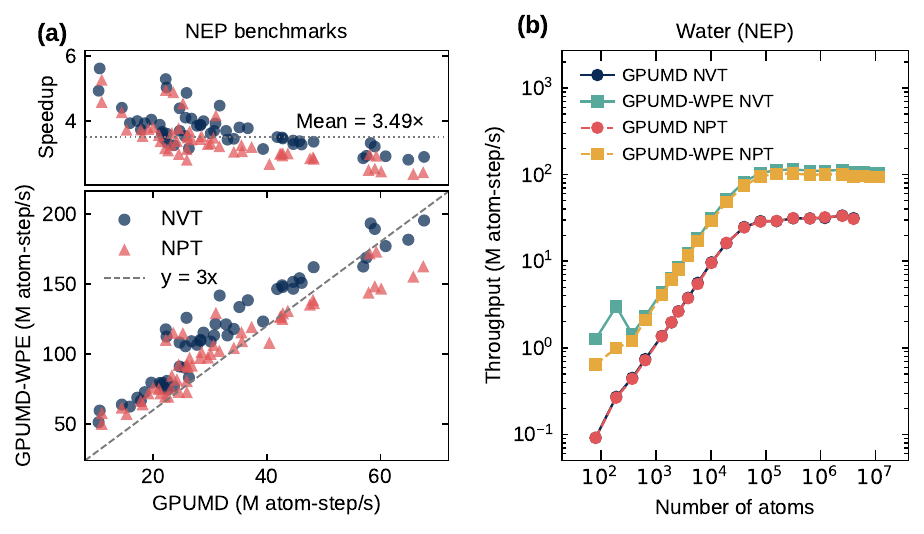}
    \caption{Ordinary NEP performance and system-size scaling on a single NVIDIA RTX 5090.
    (a) GPUMD-WPE speedups and throughput comparisons with GPUMD for 53 model-system cases under NVT and NPT (106 workloads).
    (b) Throughput scaling of GPUMD and GPUMD-WPE with system size for a fixed water NEP model under NVT and NPT. In separate memory-capacity tests, GPUMD runs out of GPU memory at 6,001,128 atoms. Accordingly, GPUMD throughput is reported for 38 of the 44 workloads shown.}
    \label{fig:nep}
\end{figure}

Across the 106 ordinary NEP workloads, \gpumdwpe{}
achieves a median throughput of 109 million atom-step/s and accelerates every workload,
with individual speedups ranging from 2.36$\times$ to 5.62$\times$,
as shown in \Cref{fig:nep}(a).
When grouped by ensemble, the geometric-mean speedups are 3.69$\times$
for NVT and 3.31$\times$ for NPT.
NEP throughput reaches 195 million atom-step/s for a 404,352-atom PbTe system under NVT.

The 53 model-system cases combine 27 initial structures and 21 potential files, spanning metals and alloys, oxides, ionic compounds, water, and layered materials.
The potentials include the general-purpose NEP89 \cite{Liang2026NEP89} and UNEP-v1 \cite{Song2024UNEP16} models, as well as models from GPUMD and PIMD studies \cite{Fan2022GPUMD,Ying2025PIMD}.
Additional models and configurations are drawn from published studies of carbon \cite{Fan2024Carbon}, water \cite{Xu2023Water,Xu2025NEPMBpol}, PbTe and Si \cite{Wu2024ForceError}, oxides \cite{Zhang2023HfO2,Yang2025AtomicArmor}, ionic frameworks \cite{Li2025ICOF}, and graphene-hBN heterostructures \cite{Liang2026GrHBN}, together with companion qNEP resources \cite{Fan2026qNEP}.
These workloads span a range of chemical compositions and NEP model shapes.

We examine system-size scaling using a fixed water NEP model under NVT and NPT, with 81 to 11,490,336 atoms, as shown in \Cref{fig:nep}(b).
For the workloads measured with both implementations, \gpumdwpe{} achieves a geometric-mean speedup of 3.55$\times$.
At the largest paired system size of 3,989,088 atoms, GPUMD-WPE reaches throughputs of 105.8 and 94.6 million atom-step/s for NVT and NPT, respectively, compared with 31.4 and 30.8 million atom-step/s for GPUMD.
These speedups of 3.37$\times$ and 3.07$\times$ show that the gains persist at the largest paired scale evaluated.
At 11,490,336 atoms, GPUMD-WPE maintains throughputs of 105.0 and 94.5 million atom-step/s under NVT and NPT, respectively.

\FloatBarrier
\subsection{qNEP and Classical Potentials}
\label{sec:benchmark:qnep-classical}

\begin{figure}[!htb]
    \centering
    \includegraphics[width=0.90\linewidth]{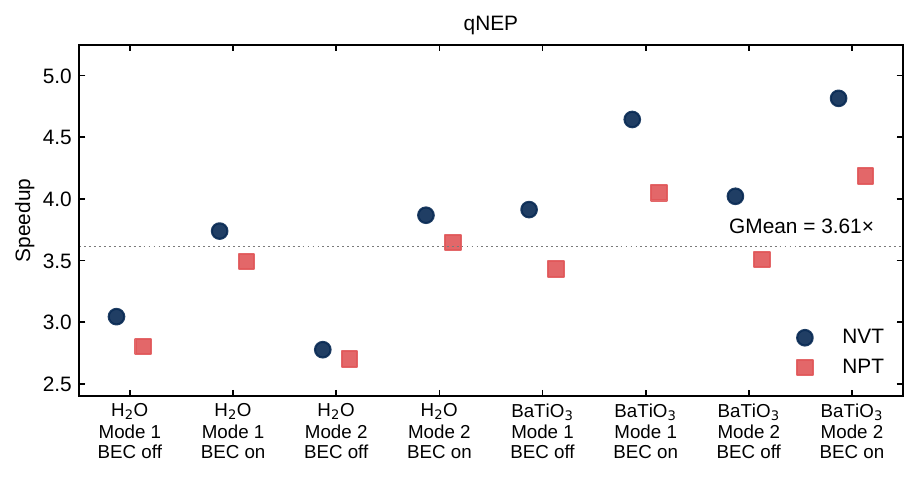}
    \caption{GPUMD-WPE speedups relative to GPUMD 5.9 for 16 qNEP workloads involving water and BaTiO$_3$, two charge modes, with and without BEC coupling, under NVT and NPT.}
    \label{fig:qnep}
\end{figure}

The \gls{qnep} benchmarks cover water and BaTiO$_3$ under NVT and NPT, using two charge modes  with and without \gls{bec} coupling \cite{Fan2026qNEP}.
Across the 16 workloads, \gpumdwpe{} achieves speedups ranging from 2.70$\times$ to 4.82$\times$, as shown in \Cref{fig:qnep}.
The BEC-disabled and BEC-enabled subsets yield geometric-mean speedups of 3.24$\times$ and 4.03$\times$, respectively.
These results characterize the complete qNEP execution paths and do not isolate the contribution of BEC evaluation.

\begin{figure}[!htb]
    \centering
    \includegraphics[width=0.90\linewidth]{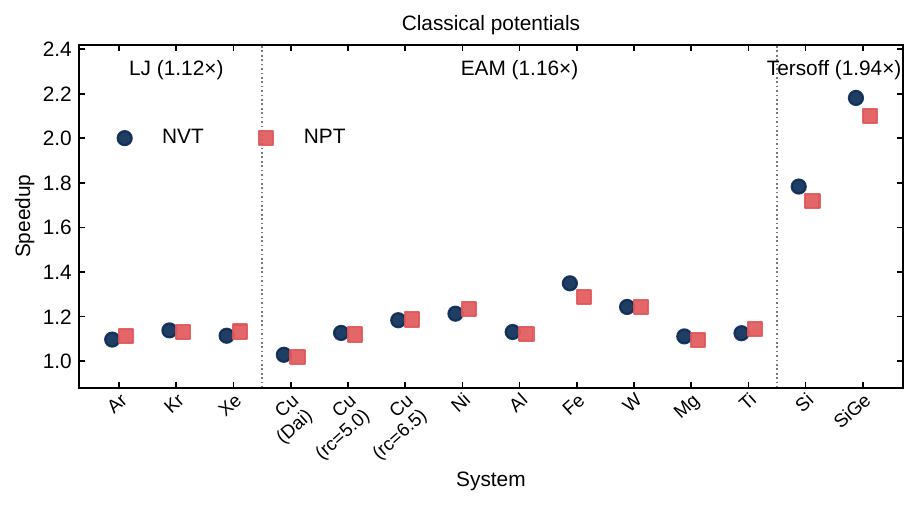}
    \caption{GPUMD-WPE speedups relative to GPUMD 5.9 for 28 classical-potential workloads covering LJ, EAM, and Tersoff-1989 under NVT and NPT.}
    \label{fig:classical}
\end{figure}

We next benchmark LJ, EAM, and Tersoff-1989 potentials (\Cref{fig:classical}).
The LJ subset comprises 442,368-atom face-centered-cubic (FCC) crystals of Ar, Kr, and Xe at 70, 95, and 130~K, respectively, using the Lennard-Jones interaction \cite{Jones1924LJ}.
The Ar benchmark is based on the public GPUMD Ar\_10A example \cite{Fan2022GPUMD}.
Across the six NVT/NPT workloads, \gpumdwpe{} achieves a geometric-mean speedup of 1.12$\times$.

The EAM subset covers FCC Cu, Ni, and Al, body-centered-cubic (BCC) Fe and W, and hexagonal close-packed (HCP) Mg and Ti using the embedded-atom method \cite{Daw1984EAM}.
Most cases employ the Zhou parameterizations \cite{Zhou2004EAM}, including two Cu variants with cutoff distances of 5 and 6.5~\AA.
An additional Cu case uses the extended Finnis-Sinclair potential of Dai et al. \cite{Dai2006EAM}.
Together, these form nine material-potential cases and 18 NVT/NPT workloads, yielding a geometric-mean speedup of 1.16$\times$.

The Tersoff subset includes diamond Si and zincblende SiGe, each with 512,000 atoms, under NVT and NPT using the Tersoff-1989 potential \cite{Tersoff1989}.
Across the four workloads, \gpumdwpe{} achieves a geometric-mean speedup of 1.94$\times$.
\gpumdwpe{} accelerates all 28 classical-potential workloads, with larger gains for Tersoff than for LJ and EAM.

\FloatBarrier
\subsection{PIMD Performance}
\label{sec:benchmark:pimd}

\begin{figure}[!htb]
    \centering
    \includegraphics[width=0.95\linewidth]{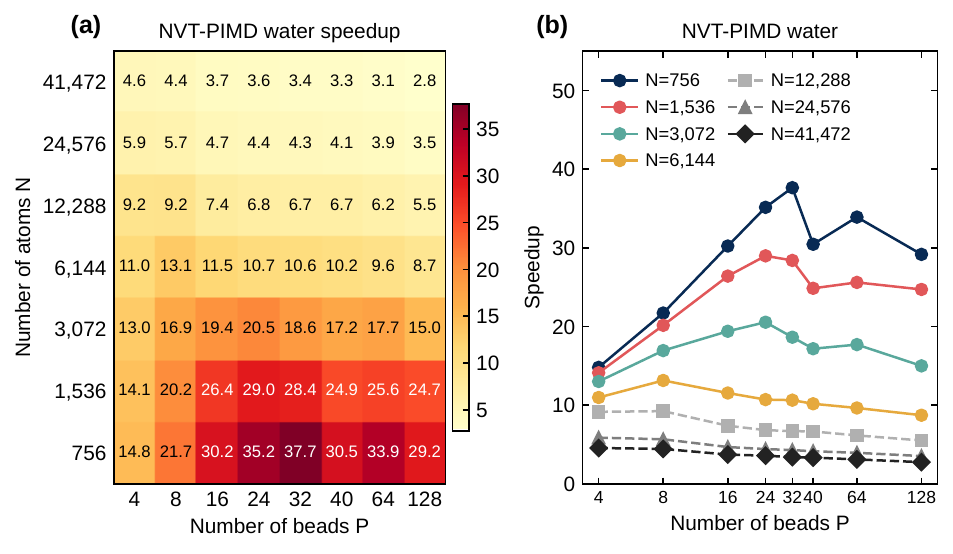}
    \caption{GPUMD-WPE speedups relative to GPUMD 5.9 for 56 water NVT-PIMD workloads spanning physical system size $N$ and bead count $P$.}
    \label{fig:pimd}
\end{figure}

We also benchmark \gpumdwpe{} for \gls{pimd}, where multiple beads require repeated potential evaluations.
The benchmark comprises 56 NVT-PIMD water workloads, combining seven physical system sizes ($N=756$--41,472 atoms) and eight bead counts ($P=4$--128), using the water model and simulation workflow from the published GPUMD PIMD study \cite{Ying2025PIMD}.
GPUMD-WPE accelerates all configurations, with speedups ranging from 2.75$\times$ to 37.7$\times$, as shown in \Cref{fig:pimd}.

The performance gains depend strongly on physical system size.
Taking the geometric mean over bead counts at each size, the speedup decreases from 28.1$\times$ at $N=756$ to 3.56$\times$ at $N=41{,}472$, while remaining between 2.75$\times$ and 4.55$\times$ at the largest system size.
This trend is consistent with a greater relative benefit from bead batching in smaller systems, where individual bead evaluations offer limited GPU parallelism.
At fixed $N$, speedup does not increase monotonically with bead count $P$.
Because GPUMD-WPE combines specialized NEP evaluation, bead batching, and other execution optimizations, these results reflect their combined
performance and do not isolate the contribution of batching.

\FloatBarrier

\section{Validation and Large-Scale Application}
\label{sec:validation}

We compare water structural correlations, PIMD thermal expansion, and \gls{hnemd} thermal conductivity obtained with \gpumdwpe{} against GPUMD results, experimental data, and published computational results.
We then present a large-scale Cu shock-compression simulation.

\subsection{Structural Validation}
\label{sec:validation:structure}

\begin{figure}[!htb]
    \centering
    \includegraphics[width=0.95\linewidth]{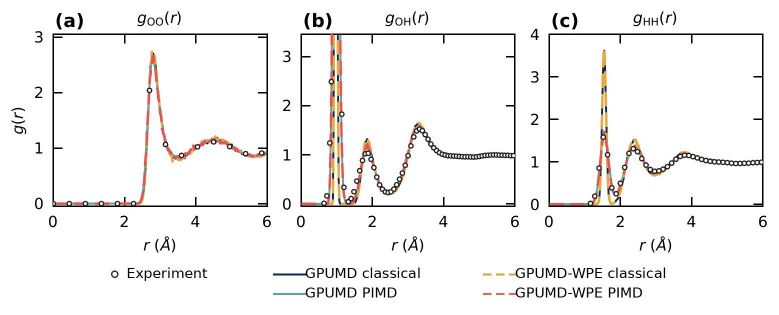}
    \caption{O-O, O-H, and H-H radial distribution functions for classical water and PIMD.
    GPUMD-WPE results are compared with those obtained using GPUMD and with experimental data.
    O-O experimental data are taken from Skinner et al. \cite{Skinner2013WaterRDF,Skinner2014Water}; the O-H/H-H experimental comparison follows Soper \cite{Soper2000Water}. The intramolecular O-H peak extends beyond the displayed vertical range.}
    \label{fig:rdf}
\end{figure}

We calculate the O-O, O-H, and H-H radial distribution functions (RDFs) of water using classical MD and PIMD with \gpumdwpe{}.
Both simulations use a 1,536-atom water system described by an NEP model \cite{Xu2023Water}, with an equilibration temperature of 300~K and a time step of 0.5~fs.
For classical MD, 25~ps of NPT equilibration is followed by 50~ps of microcanonical sampling.
For PIMD, 24 beads are employed, with 5~ps of pressure-controlled equilibration followed by 3~ps of constant-volume RDF sampling \cite{Ying2025PIMD}.
The GPUMD-WPE RDFs closely agree with those obtained from GPUMD, as shown in \Cref{fig:rdf}.
The RDFs are also broadly consistent with experimental measurements by Skinner et al. \cite{Skinner2013WaterRDF,Skinner2014Water} for O-O correlations and Soper \cite{Soper2000Water} for O-H and H-H correlations.

\FloatBarrier
\subsection{Thermophysical Validation}
\label{sec:validation:thermophysical}

\begin{figure}[!htb]
    \centering
    \includegraphics[width=0.95\linewidth]{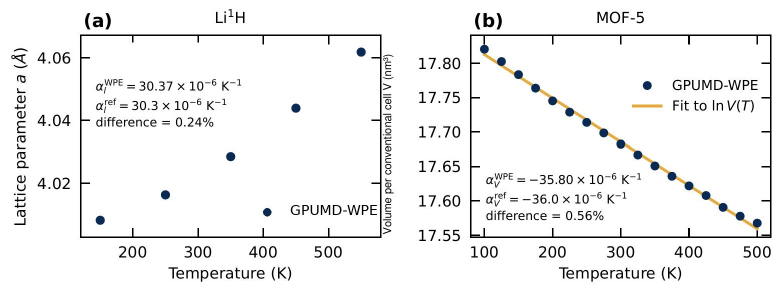}
    \caption{PIMD thermal-expansion validation with GPUMD-WPE. (a) Li$^{1}$H lattice parameter versus temperature. (b) MOF-5 volume per conventional cell versus temperature; the line is obtained by fitting $\ln V(T)$ and transforming the fit back to volume. The Li$^{1}$H estimates use 5-ps blocks. The annotations compare the calculated expansion coefficients with the corresponding published PIMD values from Ying et al. \cite{Ying2025PIMD}.}
    \label{fig:pimd-validation}
\end{figure}

We use pressure-controlled \gls{pimd} to calculate the thermal expansion coefficients in Li$^{1}$H and MOF-5 and compare them with the published PIMD results \cite{Ying2025PIMD}.
The results are shown in \Cref{fig:pimd-validation}.
At fixed pressure, the thermal expansion coefficients are 
\begin{equation}
    \alpha_l=\left(\frac{\partial\ln a}{\partial T}\right)_p, \qquad \alpha_V=\left(\frac{\partial\ln V}{\partial T}\right)_p,
\end{equation}
where $a$ is the lattice parameter and $V$ is the volume.

For Li$^{1}$H, we simulate an 8,000-atom system with 40 beads.
A centered finite-difference estimate using the lattice parameters at 250 and 350~K gives $\alpha_l=30.37\times10^{-6}$~K$^{-1}$ at 300~K, differing by only 0.24\% from the published PIMD value of $30.3\times10^{-6}$~K$^{-1}$ \cite{Ying2025PIMD}.
For MOF-5, we simulate a 27,136-atom system with 64 beads at zero pressure over 100--500~K.
The simulation-box volume is normalized by the 64 conventional cells in the $4\times4\times4$ supercell, and fitting $\ln V(T)$ yields $\alpha_V=-35.80\times10^{-6}$~K$^{-1}$, differing by 0.56\% from the published PIMD value of $-36.0\times10^{-6}$~K$^{-1}$ \cite{Ying2025PIMD}. 

\FloatBarrier
\subsection{Thermal-Transport Validation}
\label{sec:validation:thermal-transport}

We employ the \gls{hnemd} method \cite{Fan2019HNEMD}, in which an external driving field induces a nonequilibrium heat current related to the thermal conductivity tensor through
\begin{equation}
    \frac{\langle J^\mu \rangle_{\mathrm{ne}}}{TV}
    = \sum_{\nu}\kappa^{\mu\nu}F_{\mathrm e}^{\nu},
\end{equation}
where $\langle J^\mu \rangle_{\mathrm{ne}}$ is the nonequilibrium average heat current, $T$ is the temperature, $V$ is the system volume, $\kappa^{\mu\nu}$ is the thermal conductivity tensor, and $F_{\mathrm e}^{\nu}$ is the external driving-field component.
The method involves force and virial evaluations throughout the simulation.
We consider two representative systems covering different potential families: monolayer molybdenum disulfide (MoS$_2$) modeled with a fine-tuned NEP89 potential and bulk FCC copper (Cu) modeled with the Zhou EAM potential.

\begin{figure}[!htb]
    \centering
    \includegraphics[width=0.95\linewidth]{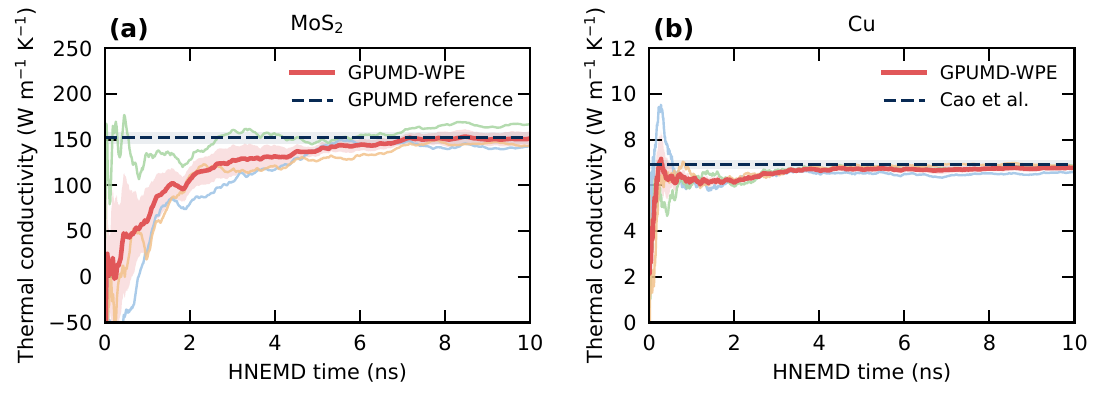}
    \caption{HNEMD lattice thermal conductivity for (a) monolayer MoS$_2$ with fine-tuned NEP89 and (b) bulk Cu with the Zhou EAM potential. Thin curves represent three independent GPUMD-WPE runs, while the red curve and shaded band show their mean and pointwise standard error, respectively. Reference values are taken from the public GPUMD tutorial for MoS$_2$ \cite{GPUMDTutorials2026} and Cao et al. for Cu \cite{Cao2025Metals}.}
    \label{fig:hnemd}
\end{figure}

For MoS$_2$, an 11,484-atom system is equilibrated at 300~K for 2~ns, followed by 10~ns of HNEMD production sampling with a time step of 1~fs and a driving field of $F_{\mathrm e}^{y}=10^{-5}$~\AA$^{-1}$.
For the monolayer system, the volume used to normalize the heat current is defined as $V=A h_{\mathrm{eff}}$, where $A$ is the in-plane area and $h_{\mathrm{eff}}=6.15$~\AA{} is the effective thickness, consistent with the reference model \cite{GPUMDTutorials2026}.
Three GPUMD-WPE runs yield a lattice thermal conductivity of $151.2\pm7.8$~W~m$^{-1}$~K$^{-1}$, as shown in \Cref{fig:hnemd}(a), in agreement with the public GPUMD tutorial result of $152.1\pm6.7$~W~m$^{-1}$~K$^{-1}$ obtained from ten runs \cite{GPUMDTutorials2026}.
For Cu, a 32,000-atom system is equilibrated at 300~K and zero pressure for 1~ns, followed by 10~ns of HNEMD production sampling with a time step of 4~fs and a driving field of $F_{\mathrm e}^{x}=10^{-4}$~\AA$^{-1}$.
Both production simulations employ the NVT ensemble with a Nos\'e-Hoover chain thermostat.
The three GPUMD-WPE runs for Cu yield a lattice thermal conductivity of $6.76\pm0.08$~W~m$^{-1}$~K$^{-1}$, as shown in \Cref{fig:hnemd}(b), consistent with the reported value of $6.9\pm0.2$~W~m$^{-1}$~K$^{-1}$ from Cao et al. \cite{Cao2025Metals}.

\FloatBarrier
\subsection{Large-Scale Shock Compression in Cu}
\label{sec:validation:cu-shock}

\begin{figure}[!htb]
    \centering
    \includegraphics[width=0.95\linewidth]{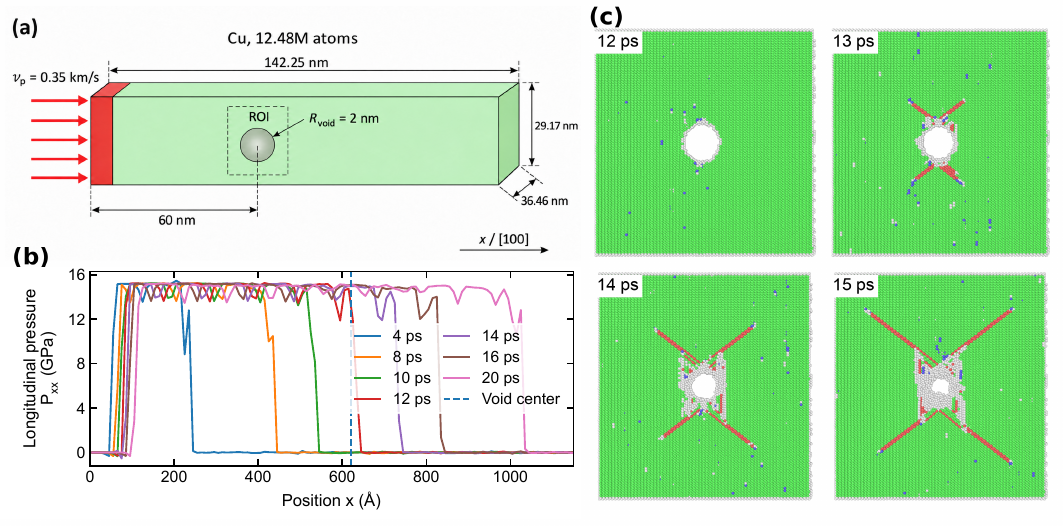}
    \caption{Large-scale Cu shock-compression simulation using GPUMD-WPE with the UNEP-v1 potential.
    (a) Schematic of the approximately 12.5-million-atom Cu specimen, showing the $[100]$ shock-propagation direction, piston velocity $v_{\mathrm p}=0.35$~km/s, and spherical void of radius 2~nm.
    (b) Longitudinal pressure profiles at selected times, with the dashed vertical line marking the initial void-center position.
    (c) Local atomic configurations around the void at 12, 13, 14, and 15~ps, colored according to adaptive common-neighbor analysis: green, FCC; red, HCP-like; blue, BCC-like; and white, other or unclassified environments.}
    \label{fig:cu-shock}
\end{figure}

We simulate shock compression of FCC Cu using the UNEP-v1 potential \cite{Song2024UNEP16}.
The specimen contains 12,477,247 atoms after introducing a spherical void of radius 2~nm and measures approximately $142.25\times36.46\times29.17$~nm$^3$.
The void is centered about 60~nm from the loaded end, with shock propagation along the $[100]$ direction.
A moving piston with a velocity of $v_{\mathrm p}=0.35$~km/s drives the shock.
Following preparation and relaxation, the system is evolved for 20~ps using a time step of 0.5~fs (40,000 steps).
Longitudinal pressure profiles are sampled every 0.1~ps using 1-nm spatial bins.
For local structural analysis, we save 400 snapshots at 0.05-ps intervals of an atom group initially selected around the void, as shown in \Cref{fig:cu-shock}(c).

\FloatBarrier

The longitudinal pressure profiles show an advancing compression front, with a compressed region reaching approximately 15~GPa behind the front, as shown in \Cref{fig:cu-shock}(b).
As the shock approaches the void, adaptive common-neighbor analysis \cite{Stukowski2012ACNA} reveals the formation and outward extension of HCP-like atomic layers between 12 and 15~ps, as shown in \Cref{fig:cu-shock}(c).
The complete 40,000-step simulation achieves a throughput of 86.2 million atom-step/s on a single NVIDIA RTX 5090.
Unlike streamlined benchmarks, this production simulation includes spatially resolved thermodynamic calculations and atomic trajectory output, while shock-induced compression increases the local atomic density and can increase the number of neighbors involved in potential evaluation.

\section{Conclusion}
\label{sec:conclusion}

We have presented \gls{wpe}, a GPU potential-evaluation engine that specializes repeated evaluations according to the loaded model and simulation-stage requirements while preserving GPUMD's general-purpose simulation workflow.
On a single NVIDIA RTX 5090, \gpumdwpe{} achieves a median
throughput of 109 million atom-step/s across 106 ordinary NEP workloads and sustains 142
million atom-step/s for a 15.1-million-atom diamond Si system. The geometric-mean speedups over GPUMD v5.9 are 3.49$\times$, 3.61$\times$, 1.24$\times$, and 10.5$\times$ for the ordinary \gls{nep}, \gls{qnep}, classical-potential, and \gls{pimd} benchmark suites, respectively.
Representative structural, thermal-transport, and thermophysical calculations show close agreement with corresponding GPUMD, experimental, or published simulation results.
A 12.5-million-atom Cu shock-compression simulation using UNEP-v1 \cite{Song2024UNEP16} further demonstrates the use of \gpumdwpe{} for large-scale nonequilibrium dynamics and the analysis of defect evolution around the void.
These results support a broader design principle for high-performance scientific software: general-purpose flexibility can coexist with workload-specialized execution.

\section*{Code and Data Availability}
\label{sec:availability}

GPUMD v5.9 is available from the GPUMD repository (\url{https://github.com/brucefan1983/GPUMD}).
WPE is available from the official WPE repository (\url{https://github.com/Wisevolve/Wisevolve-Potential-Engine}).
The WPE repository provides the C application programming interface, GPUMD adapter, benchmark inputs, and the WPE 1.0.0 shared library.

\bibliography{refs}
\end{document}